\documentclass[final,authoryear,5p,times,twocolumn]{elsarticle}

\usepackage{amssymb}
 \usepackage[switch,pagewise,running]{lineno}

\usepackage{hyperref}
\hypersetup{
    bookmarks=true,         % show bookmarks bar?
    unicode=false,          % non-Latin characters in Acrobat's bookmarks
    pdftoolbar=true,        % show Acrobat's toolbar?
    pdfmenubar=true,        % show Acrobat's menu?
    pdffitwindow=false,     % page fit to window when opened
    pdftitle={The Resolved Asteroid Program - I: (52)\,Europa},
    pdfauthor={Benoit Carry},
    pdfsubject={Planetary Science},
    pdfkeywords={},         % list of keywords
    pdfnewwindow=true,      % links in new window
    colorlinks=true,        % false: boxed links; true: colored links
    linkcolor=gray,         % color of internal links
    citecolor=blue,         % color of links to bibliography
    filecolor=gray,         % color of file links
    urlcolor=gray           % color of external links
}

\bibpunct{(}{)}{;}{a}{}{,}
\newcommand{\degr}{\ensuremath{^\circ}}
\newcommand{\arcsec}{\mbox{\ensuremath{^{\prime\prime}}}}
\newcommand{\micron}{\ensuremath{\mu}m}

\newcommand{\eg}{\textsl{e.g.}}
\newcommand{\ie}{\textsl{i.e.}}

\journal{Icarus}

\usepackage{pstricks}
\usepackage{ulem}

\graphicspath{{figures/}}

\renewcommand{\sout}[1]{\,}
\newcommand{\addr}[1]{#1}

\begin{document}

\begin{frontmatter}



\title{The Resolved Asteroid Program - Size, shape, and pole of (52)\,Europa\tnoteref{1}}
\tnotetext[1]{Based on observations
  at the W. M. Keck Observatory, which is operated as a scientific
  partnership among the California Institute of Technology, the
  University of California and the National Aeronautics and Space
  Administration. The Observatory
  was made possible by the generous financial support of the W. M. Keck
  Foundation.}

%\author{}
%\address{}

\author[swri]{W.~J.~Merline}
\ead{merline@boulder.swri.edu}
\author[aflr]{J.~D.~Drummond}
\author[imcce,esa]{B.~Carry}
\author[mpia,keck]{A.~Conrad}
\author[swri]{P.~M.~Tamblyn}
\author[eso]{C.~Dumas}

\author[tut]{M.~Kaasalainen}
\author[dlr]{A.~Erikson}
\author[dlr]{S.~Mottola}
\author[praha]{J.~{\v D}urech}
\author[cdr]{G.~Rousseau}
\author[cdr,geneve]{R.~Behrend}
\author[cdr,eurac]{G.~B.~Casalnuovo}
\author[cdr,eurac]{B.~Chinaglia}
\author[gemini]{J.~C.~Christou}
\author[swri]{C.~R.~Chapman}
\author[keck]{C.~Neyman}

\address[swri]{Southwest Research Institute, 1050 Walnut St. \# 300, Boulder, CO  80302, USA}
\address[aflr]{Starfire Optical Range, Directed Energy Directorate, Air Force Research Laboratory, Kirtland AFB, NM 87117-577, USA}
\address[imcce]{IMCCE, Observatoire de Paris, CNRS, 77 av. Denfert Rochereau, 75014 Paris, France}
\address[esa]{European Space Astronomy Centre, ESA, P.O. Box 78, 28691 Villanueva de la Ca\~{n}ada, Madrid, Spain}
\address[mpia]{Max-Planck-Institut f{\" u}r Astronomie, K{\" o}nigstuhl, 17, Heidelberg, Germany}
\address[keck]{W.M. Keck Observatory, 65-1120 Mamalahoa Highway, Kamuela, HI 96743, USA}

\address[eso]{ESO, Alonso de C\'{o}rdova 3107, Vitacura, Casilla 19001, Santiago de Chile, Chile}

\address[tut]{Tampere University of Technology, P.O. Box 553, 33101 Tampere, Finland}
\address[dlr]{Institute of Planetary Research, DLR, Rutherfordstrasse 2, 12489, Berlin, Germany}
\address[praha]{Astronomical Institute, Faculty of Mathematics and
  Physics, Charles University in Prague, V Hole\v{s}ovi\v{c}k\'ach 2,
  18000 Prague, Czech Republic}
\address[cdr]{CdR \& CdL group: Lightcurves of minor planets and variable stars}
\address[geneve]{Geneva Observatory, 1290 Sauverny, Switzerland}
\address[eurac]{Eurac Observatory, Bolzano}
\address[gemini]{Gemini Observatory, 670 N. A’ohoku Place, Hilo, Hawaii, 96720, USA}

%%%%%%%%%%%%%%%%%%%%%%%%%%%%%%%%%%%%%%%%%%%%%%%%%%%%%%%%%%%
%%%%%    TAG     %%%%%%%%%%%        ABSTRACT     %%%%%%%%%%
%%%%%%%%%%%%%%%%%%%%%%%%%%%%%%%%%%%%%%%%%%%%%%%%%%%%%%%%%%%
\begin{abstract}
  With the adaptive optics (AO) system on the 10\,m Keck-II telescope,
  we acquired a high quality set of 84 images at 14 epochs of asteroid
  (52)\,Europa on 2005\,January\,20,
  when it was near opposition. The
  epochs covered its 5.63\,h rotation period and, by following its changing
  shape and orientation on the plane of sky,
  we obtained its triaxial ellipsoid dimensions
  and spin pole location.
  An independent determination from images at
  three epochs obtained in 2007 is in good agreement with these
  results. By combining these two data sets, along with a single epoch
  data set obtained in 2003,
  we have derived a global
  fit for (52)\,Europa of diameters
  $a \times b \times c = ( 379 \times 330 \times 249 ) \pm (16 \times 8 \times  10)$\,km,
  yielding a volume-equivalent spherical-diameter of
  $\sqrt[3]{abc} = \addr{315} \pm \addr{7}$\,km,
  and a prograde rotational pole
  within \sout{3}\addr{7}\degr~of [RA; Dec] = [257\degr; +12\degr] in
  an Equatorial J2000 reference frame (Ecliptic:  255\degr; +35\degr).
  Using the average of all mass determinations available for
  (52)\,Europa, we derive a density of 1.5\,$\pm$\,0.4\,g\,cm$^{-3}$,
  typical of C-type asteroids.
  Comparing our images with the shape model
  of Michalowski et al. (Astron. and Astrophys. 416, p 353, 2004),
  derived from optical lightcurves, illustrates excellent
  agreement, although several edge features visible in the images are not
  rendered by the model.
  We therefore derived a complete 3-D description of (52)\,Europa's
  shape using the KOALA algorithm by combining our 18 AO imaging
  epochs with 4 stellar occultations and 49 lightcurves.
  We use this 3-D shape model to assess these
  departures from ellipsoidal shape.
\addr{Flat facets (possible giant craters) appear to be
less distinct on (52)\,Europa than on other C-types that have
been imaged in detail, (253)\,Mathilde and (511)\,Davida.}  
\sout{Europa shows
  less-pronounced facets (possible giant craters)
  than other C-types, such as Mathilde or Davida, but
  we show that this}
\addr{We show that fewer giant craters, or smaller craters, }
is consistent with its
  expected impact history. 
  Overall, asteroid (52)\,Europa is still
  well modeled as a smooth triaxial
  ellipsoid with dimensions constrained by observations obtained over
  several apparitions. 
\end{abstract}


\end{frontmatter}

% \linenumbers

 \pagewiselinenumbers

%%%%%    TAG     %%%%%%%%%%%    INTRODUCTION     %%%%%%%%%%
%%%%%%%%%%%%%%%%%%%%%%%%%%%%%%%%%%%%%%%%%%%%%%%%%%%%%%%%%%%
\section{Introduction}

\begin{figure*}[!t]
\begin{center}
  \includegraphics[width=.8\textwidth]{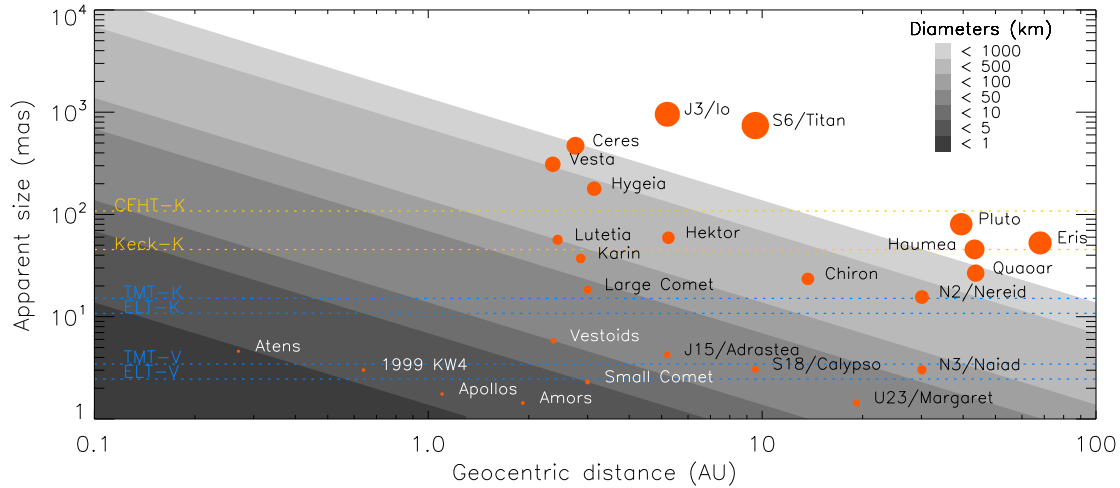}
  \caption[Asteroid observable with adaptive-optics imaging]{%
    Apparent angular sizes of Solar
    System objects. Asteroid, moon, comet, and TNO diameters are
    plotted against their geocentric 
    distances, defined as the difference between their semi-major axis and 1\,AU.
    Symbol size corresponds to physical diameter.
    Gray scales represent the changing apparent size with geocentric distance.
    A body of a given size moves along the oblique lines as its distance from the Earth changes.
    The angular resolutions at CFHT, Keck and future TMT and E-ELT are also
    shown for different filters
    (V: 0.6\,\micron, and K: 2.2\,\micron).
    Typical NEA populations (Apollos, Atens, and Amors) are also
    shown, as represented by (1566)\,Icarus,
    (99942)\,Apophis, and (433)\,Eros, respectively.
  \label{fig: RAP}
  }
\end{center}
\end{figure*}

  \indent Direct, accurate measurements of asteroid shapes, sizes, and
  pole positions are now possible for larger asteroids
  that can be spatially resolved using
  the Hubble Space Telescope (HST) or
  large ground-based telescopes equipped with adaptive optics (AO).
  Physical and statistical
  study of asteroids requires accurate knowledge of these parameters. Improved
  sizes permit improved estimates of albedo, in turn allowing better interpretation
  of surface composition. In those cases where we have an estimate of
  the mass, for example from the presence of a satellite, the uncertainty in the
  volume of the asteroid is the overwhelming uncertainty in attempts to derive
  its density
  \citep{2002-AsteroidsIII-2.2-Merline}. Of course, density is the
  single most critical
  observable having a bearing on bulk composition, porosity, and
  internal structure
  \citep{2002-AsteroidsIII-2.2-Merline, 2002-AsteroidsIII-4.2-Britt,
    2006-LPI-37-Britt}.
  With our technique of determining the size of an asteroid by
  following its changing apparent size, shape, and orientation, the
  uncertainties in
  volume can now be reduced to the level of the mass uncertainty, vastly
  improving
  our confidence in the derived asteroid densities.
  The improvement
  comes about
  because we can see the detailed shape, track edge or surface
  features during rotation, and often can make an immediate pole
  determination.\\ 
  \indent Dedicated study of asteroids
  now allows directly observable
  shape profiles, and already has shown that some asteroids show
  large departures from a reference ellipsoid that may provide clues
  to the body's
  response to large impacts over time
  \citep[\eg, (4)\,Vesta,][]{1997-Science-277-Thomas}.
  For asteroid (511)\,Davida, we suggested
  \citep{2007-Icarus-191-Conrad}
  that such
  features (\eg, large flat facets)
  may be analogs of the giant craters,
  seen edge-on, in the images of (253)\,Mathilde
  during the NEAR mission
  \citep{1999-Icarus-140-Veverka} flyby.
  If giant craters are evident on these
  surfaces, they can be related to the impact history and impact
  flux over time, and there is
  some chance they can be associated with asteroid families or clusters that are
  being identified by numerical back-integration
  and clustering of orbital elements
  \citep[\eg,][]{2002-Nature-417-Nesvorny}.\\
  \indent As we have demonstrated with asteroid (511)\,Davida
  \citep{2007-Icarus-191-Conrad},
  we can derive an asteroid's triaxial ellipsoid dimensions and rotational pole
  location in a single night. However, we now have developed the ability to
  combine sets of similar observations obtained at different viewing aspects to make a
  global fit to all of the images, drastically reducing dimension
  uncertainties that might be due to sparse rotational sampling or
  peculiar observing geometries
  (Drummond et al., in preparation).
  The leverage of widely spaced
  observations and the accompanying range of viewing aspects allows unprecedented
  accuracy in derived parameters. We can then use these estimates to
  project the apparent size and shape of an asteroid into the past or future,
  making the asteroid useful as a reference or calibration object.\\
  \indent Here we report on the physical properties of the
  asteroid (52)\,Europa as a part of our on-going
  \textsl{Resolved Asteroid Program}.
  We routinely image the apparent disk of asteroids, and search
  their close vicinity for companions,
  aiming at setting better
  constraints on their spin-vector coordinates, 3-D shapes, sizes,
  and multiplicity.
  One of our main goals is to derive (or
  better constrain) 
  their densities.
  We use two independent methods to determine size,
  shape, and pole position of the target asteroids. One of
  these is based on the assumption that the
  shape is well-described by a smooth triaxial ellipsoid
  \citep[see][for instance]{2000-LGSAO-Drummond,
    2009-Icarus-202-Drummond, 2010-AA-523-Drummond}.
  Our other method allows construction of
  full 3-D shape models by combining our AO images with other
  data types, when available
  \citep[\eg, optical lightcurves and stellar occultations,
    see][]{2010-Icarus-205-Carry-a, 2010-AA-523-Carry}, in
  the technique we call KOALA 
  \citep[Knitted Occultation, Adaptive-optics and Lightcurve
      Analysis, see][]{2010-Icarus-205-Carry-a, 2011-IPI-5-Kaasalainen}.\\
  \indent The best angular resolution,
  approximated by
  $\theta$\,=\,$\lambda$/$\mathcal{D}$ (radian), with $\lambda$ the
  wavelength and $\mathcal{D}$ the diameter of the telescope
  aperture, of current ground-based optical telescopes is
  about 0.04\arcsec~(Keck/NIR).
  Due to systematics, however, we
  have found that our ability to accurately measure sizes and
  details of the apparent shape
  degrades below about 0.10\arcsec, based on simulations and
  observations of the moons of Saturn and
  simulations
  \citep{these-carry, 2009-DPS-41-Drummond}.
  The sample of observable asteroids
  (\ie, having angular sizes that get above about 0.10\arcsec)
  is therefore limited to about 200.\\
  \indent This limit in angular resolution can be converted to a
  physical diameter. As can be seen in Fig.~\ref{fig: RAP}, we can probe the size
  distribution of main-belt asteroids down to about 100\,km, while
  Pluto is the only Trans-Neptunian Object (TNO) whose apparent disk can be
  resolved.
  At opportune times, we have been able to resolve the disks of
  Near-Earth Asteroids
  \citep[NEAs, for example, see][]{2011-IAUC-9242-Merline, 2012-ACM-Merline}.
  The next generation of optical facilities will allow an improvement in
  angular resolution by a factor of
  3-4 due to mirror size alone (30\,m for TMT and 40\,m for E-ELT), allowing
  the observation of more than 500 asteroids, even if we
  consider only objects that reach half (or 0.05\arcsec) of the current
  size limits (We computed the expected
  apparent diameter of asteroids for the 2020--2030 period, and
  counted objects when apparent diameters reach 0.05\arcsec~within this
  period.)
  Second-generation instruments with high-Strehl AO corrections into
  visible wavelengths are planned for these large ground-based telescopes, providing
  another factor of 5 improvement due to operation at shorter wavelengths. Together, 
  these two factors should provide more than 
  an order-of-magnitude improvement with respect to current
  resolution. Almost 7\,000 asteroids should then be observable with
  apparent diameters greater than 0.01\arcsec.
  This breakthrough in
  imaging capabilities will also enable the spatial resolution of
  apparent disks of TNOs larger than 500\,km,
  larger moons ($\sim$100\,km) of Uranus and Neptune,
  small moons of Jupiter and Saturn,
  main-belt asteroids of few tens of kilometers, and
  NEAs of several hundred meters in favorable
  conditions (Fig.~\ref{fig: RAP}).\\
%
%%%%%%------ FIGURE --- Begin --- RAP Targets ------%%%%%%
%%%%%%------ FIGURE ---  End  --- RAP Targets ------%%%%%%
%
%
%

%%%%%%%%%%%%%%%%%%%%%%%%%%%%%%%%%%%%%%%%%%%%%%%%%%%%%%%%%%%
%%%%%    TAG     %%%%%%%%%%%    OBSERVATIONS     %%%%%%%%%%
%%%%%%%%%%%%%%%%%%%%%%%%%%%%%%%%%%%%%%%%%%%%%%%%%%%%%%%%%%%
\section{Disk-resolved imaging observations\label{sec: obs}}

  \indent For asteroid (52)\,Europa, our primary data set was taken on
  2005\,January\,20. In addition, we observed (52)\,Europa at one epoch on
  2003\,October\,12, and at three epochs on 2007\,May\,28.  
  In 2005 we obtained adaptive optics images of (52)\,Europa at H
  (1.6\,\micron) and Kp
  (2.1\,\micron) bands with NIRC2
  \citep{2004-AppOpt-43-vanDam} on the Keck II 10 m telescope, and give the observing log in
  Table~\ref{tab: obs}.
  %Images were acquired at a total of 18 epochs, one
  %on 2003\,October\,12, 14 on 2005\,January\,20, and 3 on 2007\,May\,28.
  The 2003 and 2005 images were taken using the first generation Keck wave-front
  controller; the 2007 images were taken using Keck's next generation wavefront
  controller
  \citep[NGWFC,][]{2007-KAON-489-VanDam}
   under similar conditions.
  Strehl ratios were 30\%, 27\%, and 40\% on average, respectively, for the 2003,
  2005, and 2007 epochs.
  The latter, higher, value reflects the NGWFC
  changes which, in addition to a new detector, include improvements to the
  electronics and to the software. The data set consists of 111 images: 
  9 from 2003, 84 from 2005, and 18 from 2007, that result in 18
  composite images (Table~\ref{tab: obs}).\\
  \indent 
  Although less extensive and at a larger distance from Earth, the 2007 data add
  an important new epoch to our 2005 data. By combining all 3 data
  sets (2003, 2005, and 2007), our goal was to derive a global 
  fit that spans a wide range of viewing geometries and provide tight
  constraints on the size, shape, and pole for (52)\,Europa.\\
  \indent When observing at Kp in good seeing conditions, adaptive optics on Keck 
  II delivers diffraction-limited \addr{resolution} elements of width
  approximately 50\,milli-arcsecond (mas).
  We used the narrow plate-scale
  (9.942\,$\pm$\,0.050\,mas/pixel) of the NIRC2 camera,
  oriented North-up
  \citep[$\pm$0.15\degr,][]{2007-AJ-133-Konopacky}
  for all the observations.

%%%%%%%------ TABLE --- Begin --- Observational Conditions ------%%%%%%
\begin{table*}[!t]
\begin{center}
  \begin{tabular}{cccccccc}
    \hline\hline
    Date  & $\Delta$ & $r$ & $\phi$ & m$_V$ & $\varphi$ & Rotation
    phase & \addr{Filter} \\% & SEP$_\lambda$ & SEP$_\beta$ & SSP$_\lambda$ & SSP$_\beta$ \\
     (UT) & (AU) & (AU) & (\degr) & (mag) & (\arcsec) & (\degr)\\%& (\degr) & (\degr)  & (\degr) & (\degr)    \\
    \hline%
    2003-10-12 - 11:48 & 3.02 & 2.07 &  7.2 & 10.8 & 0.25 &  26 & \addr{Kp} \\%&  15 &  49 &  13 &  42 \\
    2005-01-20 - 10:39 & 2.79 & 1.84 &  5.5 & 10.3 & 0.28 &   6 & \addr{Kp} \\%&  35 &  23 &  32 &  27 \\
    2005-01-20 - 10:43 & 2.79 & 1.84 &  5.5 & 10.3 & 0.28 &   9 & \addr{H} \\%&  32 &  23 &  28 &  27 \\
    2005-01-20 - 11:25 & 2.79 & 1.84 &  5.5 & 10.3 & 0.28 &  55 & \addr{Kp} \\%& 346 &  23 & 343 &  27 \\
    2005-01-20 - 11:28 & 2.79 & 1.84 &  5.5 & 10.3 & 0.28 &  58 & \addr{H} \\%& 343 &  23 & 339 &  27 \\
    2005-01-20 - 12:02 & 2.79 & 1.84 &  5.5 & 10.3 & 0.28 &  95 & \addr{Kp} \\%& 307 &  23 & 303 &  27 \\
    2005-01-20 - 12:04 & 2.79 & 1.84 &  5.5 & 10.3 & 0.28 &  97 & \addr{H} \\%& 307 &  23 & 303 &  27 \\
    2005-01-20 - 13:01 & 2.79 & 1.84 &  5.5 & 10.3 & 0.28 & 157 & \addr{Kp} \\%& 244 &  23 & 240 &  27 \\
    2005-01-20 - 13:04 & 2.79 & 1.84 &  5.5 & 10.3 & 0.28 & 160 & \addr{H} \\%& 242 &  23 & 238 &  27 \\
    2005-01-20 - 13:45 & 2.79 & 1.84 &  5.5 & 10.3 & 0.28 & 204 & \addr{Kp} \\%& 198 &  23 & 194 &  27 \\
    2005-01-20 - 13:48 & 2.79 & 1.84 &  5.5 & 10.3 & 0.28 & 206 & \addr{H} \\%& 195 &  23 & 191 &  27 \\
    2005-01-20 - 14:16 & 2.79 & 1.84 &  5.5 & 10.3 & 0.28 & 237 & \addr{Kp} \\%& 165 &  23 & 161 &  27 \\
    2005-01-20 - 14:18 & 2.79 & 1.84 &  5.5 & 10.3 & 0.28 & 239 & \addr{H} \\%& 165 &  23 & 161 &  27 \\
    2005-01-20 - 15:02 & 2.79 & 1.84 &  5.5 & 10.3 & 0.28 & -74 & \addr{Kp} \\%& 116 &  23 & 112 &  27 \\
    2005-01-20 - 15:05 & 2.79 & 1.84 &  5.5 & 10.3 & 0.28 & -71 & \addr{H} \\%& 116 &  23 & 112 &  27 \\
    2007-05-28 - 11:44 & 3.41 & 2.69 & 13.3 & 11.9 & 0.19 & 105 & \addr{Kp} \\%& 298 & -40 & 283 & -48 \\
    2007-05-28 - 12:54 & 3.41 & 2.69 & 13.3 & 11.9 & 0.19 & 179 & \addr{Kp} \\%& 223 & -40 & 209 & -48 \\
    2007-05-28 - 13:01 & 3.41 & 2.69 & 13.3 & 11.9 & 0.19 & 186 & \addr{Kp} \\%& 216 & -40 & 201 & -48 \\
    \hline
  \end{tabular}
  \caption{
    Observing log:
    heliocentric distance ($\Delta$),
    range to observer ($r$),
    phase angle ($\phi$),
    visual apparent magnitude (m$_V$),
    angular diameter ($\varphi$),
    and arbitrary rotation phase
    (zero phase being defined for a lightcurve maximum, \ie,
    when the apparent cross-section of (52)\,Europa is the largest)
    for each epoch
    (reported in UT).
    \label{tab: obs}
  }
\end{center}
\end{table*}
%%%%%%%------ TABLE ---  End  --- Observational Conditions ------%%%%%%

%%%%%%%%%%%%%%%%%%%%%%%%%%%%%%%%%%%%%%%%%%%%%%%%%%%%%%%%%%%%
%%%%%%    TAG     %%%%%%%%%%%   TRI-AXIAL         %%%%%%%%%%
%%%%%%%%%%%%%%%%%%%%%%%%%%%%%%%%%%%%%%%%%%%%%%%%%%%%%%%%%%%%
\section{Triaxial Ellipsoid (TE) Assumption}
  \subsection{2005\,January\,20}
\begin{figure}[!ht]
\begin{center}
  \includegraphics[width=.45\textwidth]{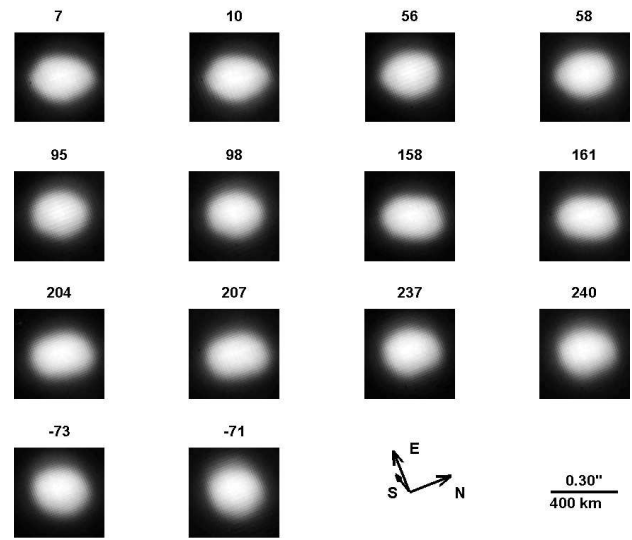}
  \caption{%
    Sky-subtracted, flat-fielded, shifted, and added,
    images of (52)\,Europa, from 2005, before deconvolution,
    rotated so that the asteroid's
    spin axis is vertical.
    Although the direction to the Sun is indicated, the solar
    phase angle was only 5.5\degr, making the Sun nearly
    perpendicular to the plane of the figure.
    The rotational phase in degrees, $\pm$ 360\degr, of each
    tile is placed on top of it 
    for placement in Fig~\ref{fig: stea05}.
    The Kp-band
    images are in the first and third columns while the H-band images
    always follow by a few degrees rotation in the second and fourth
    columns.
  \label{fig: sna}
  }
\end{center}
\end{figure}

\begin{figure}[!ht]
\begin{center}
  \includegraphics[width=.45\textwidth]{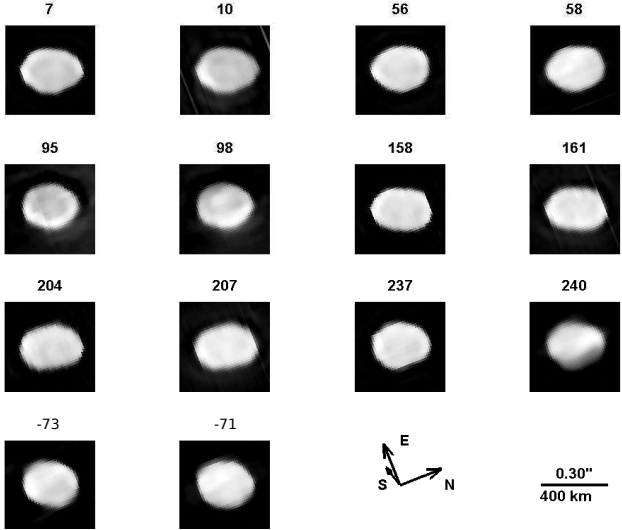}
  \caption{%
    Same as in Fig.~\ref{fig: sna}
    for the \textsc{Mistral} deconvolved images of (52)\,Europa.
% except deconvolved images of Europa.
  \label{fig: deconv}
  }
\end{center}
\end{figure}

    \indent Each of seven sets of six H-band images and seven sets of
    six Kp-band images
    of asteroid (52)\,Europa obtained in 2005 was sky-subtracted, and then fit
    in the Fourier plane for the asteroid and Lorentzian PSF, using our
    method of Parametric Blind Deconvolution
    \citep[PBD, as described by][]
      {1998-Icarus-132-Drummond, 2000-LGSAO-Drummond, 2007-Icarus-191-Conrad}.
    Asteroid ellipse parameters were computed as weighted means
    from each set of six images obtained at \addr{each} filter and each rotational
    phase or epoch.  These ellipse parameters
    (apparent major axis length $\alpha$,
    minor axis length $\beta$,
    and an orientation angle PA$_{\alpha}$),
    were then used to convert
    the series of apparent diameters and
    orientations to the full triaxial-ellipsoid diameters and
    direction of (52)\,Europa's
    rotational pole
    through a non-linear least squares inversion
    \citep[see][for instance]{2000-LGSAO-Drummond}.
    The results of the fit are given in Table~\ref{tab: stea05}.\\
%
%
%
%
%%%%%%%------ TABLE --- Begin --- STEA Parameters ------%%%%%%
\begin{table*}[!t]
\begin{center}
  \begin{tabular}{ccccc}
    \hline\hline
    Parameter  & PBD  & \textsc{Mistral} & Edges & Mean \\
    \hline
    a (km)  &  377 $\pm$ 3  &  376 $\pm$ 3  &  381 $\pm$  4  &  378 $\pm$ 3 \\
    b (km)  &  331 $\pm$ 3  &  332 $\pm$ 3  &  335 $\pm$  4  &  \addr{332}\sout{333} $\pm$ 3 \\
    c (km)  &  236 $\pm$ 9  &  246 $\pm$ 8  &  249 $\pm$ 10  &  244 $\pm$ 8 \\
    \hline
    SEP$_\beta$ (\degr)& +27 $\pm$ 3 & +25 $\pm$ 3 & +25 $\pm$ 5 & +25 $\pm$ 3 \\
    PA$_\textrm{node}$ (\degr)& 339 $\pm$ 1 & 339 $\pm$ 1 & 338 $\pm$ 1 & 338 $\pm$ 1 \\
    $\psi_0$ (UT) & 10.35 $\pm$ 0.03 & 10.33 $\pm$ 0.03 & 10.28 $\pm$ 0.04 & 10.30 $\pm$ 0.03 \\
    \hline
    EQJ2000 ($\alpha_0$,$\delta_0$ in \degr) &
    261;+10 & 260;+11 & 259;+12 & 260;+11 \\
    $\sigma$ radius (\degr)  & \addr{1}\sout{2} & \addr{1}\sout{2} & \addr{1}\sout{2} & \addr{1}\sout{2} \\
    ECJ2000 ($\lambda_0$,$\beta_0$ in \degr) &
    260;+34 & 258;+34 & 257;+35 & 258;+34 \\
    \hline
  \end{tabular}
  \caption{
    Triaxial-ellipsoid parameters for our 2005 data, with three
    different data-processing methods: PBD images,
    \textsc{Mistral} deconvolved images, and edge fitting.
    The average values for the parameters are reported in the last
    column. The quantities derived from the fits of the 2005 data are:
    triaxial ellipsoid diameters $a$, $b$, and $c$;  the sub-Earth
    latitude SEP$_\beta$;
    the line of nodes (the intersection of
    the asteroid's equator and the plane of the sky) $PA_{\textrm{node}}$;
    and the UT of the instant when the long axis $a$ lies in the plane of
    the sky along the line of nodes $\psi_0$.
    Uncertainties reported here are formal error bars of the fit, see
    the text for a discussion on the systematics.
    \label{tab: stea05}
  }
\end{center}
\end{table*}
%%%%%%%------ TABLE ---  End  --- STEA Parameters ------%%%%%%
%
%
    \indent In addition to the direct PBD methodology, as
    cross-checks, we use two additional avenues to get to the
    triaxial-ellipsoid solutions. In the first of these, the data
    \sout{are} \addr{were} flat-fielded, shifted, and added at each
    rotational epoch (Fig.~\ref{fig: sna}), and
    a single deconvolved image was created with the \textsc{Mistral}
    algorithm \citep{2004-JOSAA-21-Mugnier}, for each epoch and each filter.
    These seven Kp and seven H deconvolved images
    (Fig.~\ref{fig: deconv}) were again fit in the Fourier
    plane for their apparent ellipse parameters, and the series was fit for the full
    triaxial solution, also given in Table~\ref{tab: stea05}.\\
%
%%%%%%------ FIGURE --- Begin --- SnA Images - Undeconvolved ------%%%%%%
%%%%%%------ FIGURE ---  End  --- SnA Images - Undeconvolved ------%%%%%%
%
%

%%%%%%------ FIGURE --- Begin --- SnA Images - Deconvolved ------%%%%%%
%%%%%%------ FIGURE ---  End  --- SnA Images - Deconvolved ------%%%%%%
%
%

%%%%%%------ FIGURE --- Begin --- globes ------%%%%%%
\begin{figure}[!t]
\begin{center}
  \includegraphics[width=.35\textwidth]{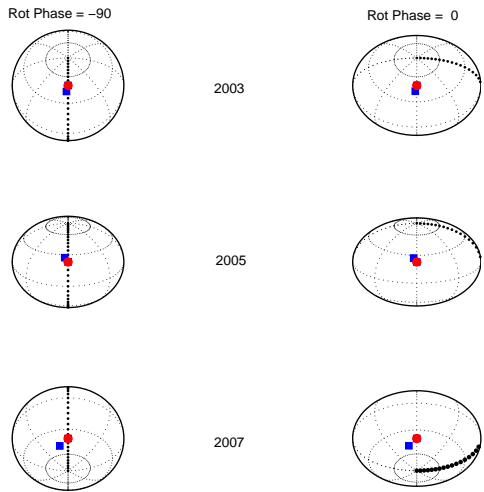}
  \caption{%
    Plane-of-sky orientation of
    (52)\,Europa as seen during the 3 observing dates
    analyzed. The grids are in equatorial
    coordinates, with north up, east left.
    The blue square is the subsolar point and the
    red circle is the sub-Earth point. Two
    views for each are shown: 
    the maximum (Rot Phase = 0) cross-section
    and the minimum 
    (Rot Phase = -90) cross-section for that
    date. These phases are the same as those
    listed in the tables and
    Figs.~\ref{fig: stea05}, \ref{fig: stea07}, and~\ref{fig: stea03}.
    The bold dotted line represents the line
    defined as longitude = 0, according to IAU
    convention
    \citep[see][]{2011-CeMDA-109-Archinal}.
    The longitude
    is related to the rotational phase by:
    longitude = 270\degr\,-\,Rot\,phase.  The sense of
    rotation is given by the right-hand rule
    here, with the (positive) pole always
    northward, and can be discerned in the
    figure from the advancement of the
    bold dotted line by 90\degr. 
  \label{fig: globes}
  }
\end{center}
\end{figure}
%%%%%%------ FIGURE ---  End  --- globes ------%%%%%%

%%%%%%------ FIGURE --- Begin --- STEA Curve ------%%%%%%
\begin{figure}[!t]
\begin{center}
  \includegraphics[width=.45\textwidth]{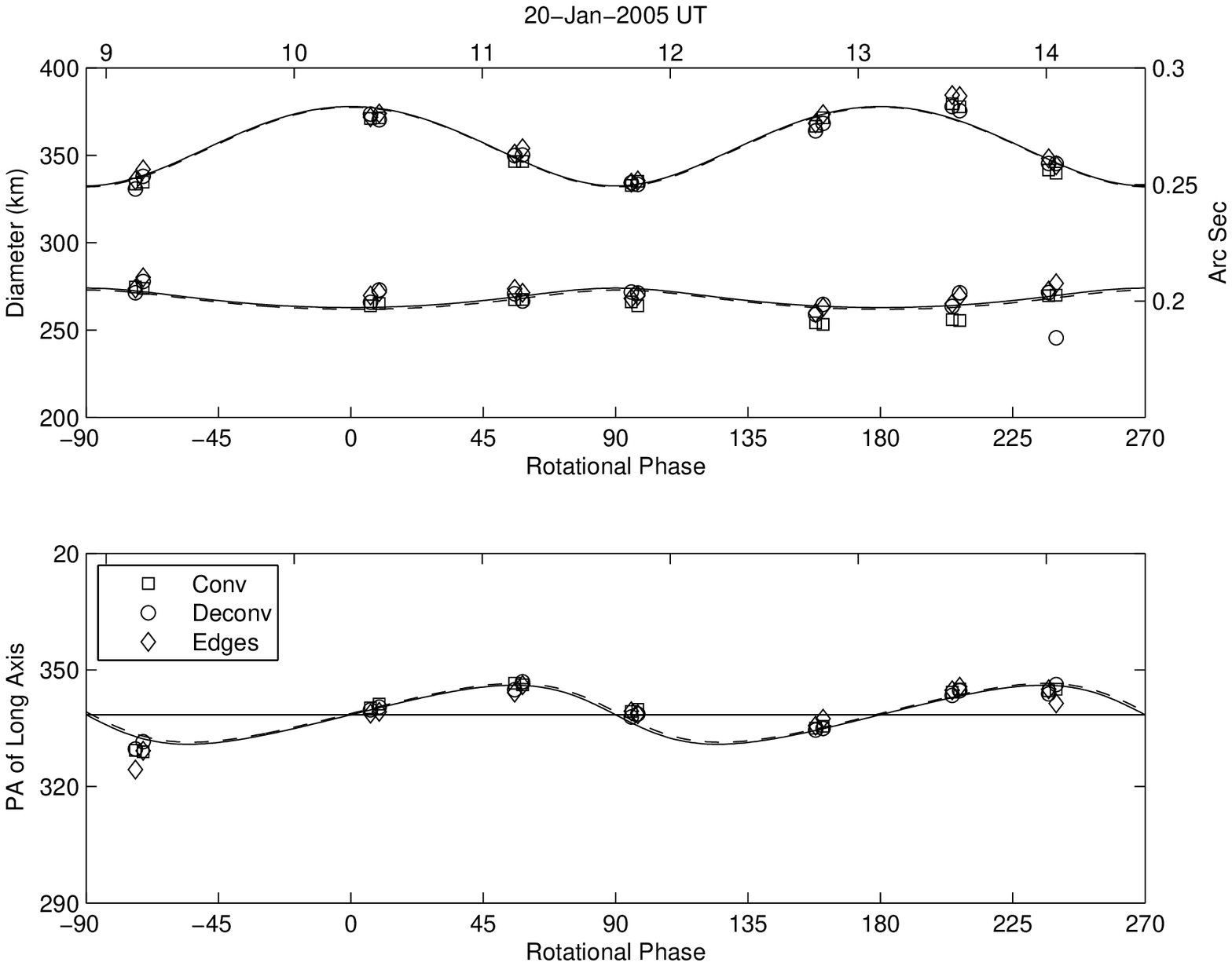}
  \caption{%
    Triaxial ellipsoid fit to measured ellipse parameters for our 2005 data.
    In the upper subplot, each image's long and short axis dimensions are
    plotted along the upper and lower lines, respectively.
    The H-band epochs follow the Kp-band epochs by a few degrees,
    and the different symbols represent
    the different methods used to extract the ellipse parameters
    (PBD or Conv, \textsc{Mistral} or Deconv, and Edges).
    The solid lines
    are the prediction for the projected (full) ellipses from the mean
    triaxial ellipsoid parameters (Table~\ref{tab: stea05}).
    The dashed lines are for the ellipse parameters for
    the terminator ellipse, which, because the solar phase angle is
    only 5.5\degr, fall
    on the solid lines.
    The data should lie approximately midway between the
    dashed and solid lines (here, that means on the coincident
    solid/dashed lines).
    The lower subplot shows the position angle of the long axis (PA$_\alpha$)
    with the same conventions.
  \label{fig: stea05}
  }
\end{center}
\end{figure}
\begin{figure}[!t]
\begin{center}
  \includegraphics[width=.35\textwidth]{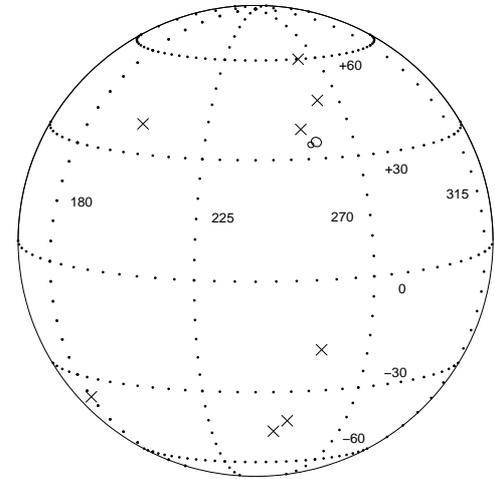}
  \caption{%
    Pole locations for (52)\,Europa on the Ecliptic globe. The two circles
    denote the uncertainty areas around the pole found for 2005
    (larger) and
    2007 (smaller), while X's show the positions found from previous
    workers using lightcurves.
  \label{fig: pole}
  }
\end{center}
\end{figure}

%%%%%%------ FIGURE ---  End  --- STEA Curve ------%%%%%%
%
%
    \indent Finally, ellipse parameters were derived from fitting the
    edges produced by a Laplacian of Gaussian wavelet transform
    \citep{2008-AA-478-Carry} on the \textsc{Mistral} deconvolved images.
    A full triaxial solution can then be found
    from these ellipse parameters, and is given in
    Table~\ref{tab: stea05}.
    \indent The adopted triaxial solution for (52)\,Europa, independently determined
    from the 2005 data, is derived from the series of mean ellipse
    parameters at each epoch, that is, from the mean of the PBD images,
    the \textsc{Mistral} deconvolved images, and
    the edge-fitting at each epoch.
    This preferred mean fit is
    plotted against observations in Fig.~\ref{fig: stea05}.
    The location of the pole on the Ecliptic
    globe is shown in Fig.~\ref{fig: pole}, along with the locations
    derived from lightcurves analysis by others.\\
    \indent Our imaging of (511)\,Davida \citep{2007-Icarus-191-Conrad}
    showed large edge features
    that could be followed during rotation, even
    in the raw images.  While there may be similar features on (52)\,Europa, they
    do not appear as consistently in the edge profiles and are not as
    easy to track. The features are not as large or prominent as those
    on Davida, relative to our reference ellipsoid. 
    Later in the paper, we use 3-D shape modeling
    to try to study these departures from a pure ellipsoid shape.

%
%%%%%%------ FIGURE --- Begin --- Pole solution ------%%%%%%
%%%%%%------ FIGURE ---  End  --- Pole solution ------%%%%%%
%
%

  \subsection{2007\,May\,28}
\begin{figure}[!t]
\begin{center}
  \includegraphics[width=.45\textwidth]{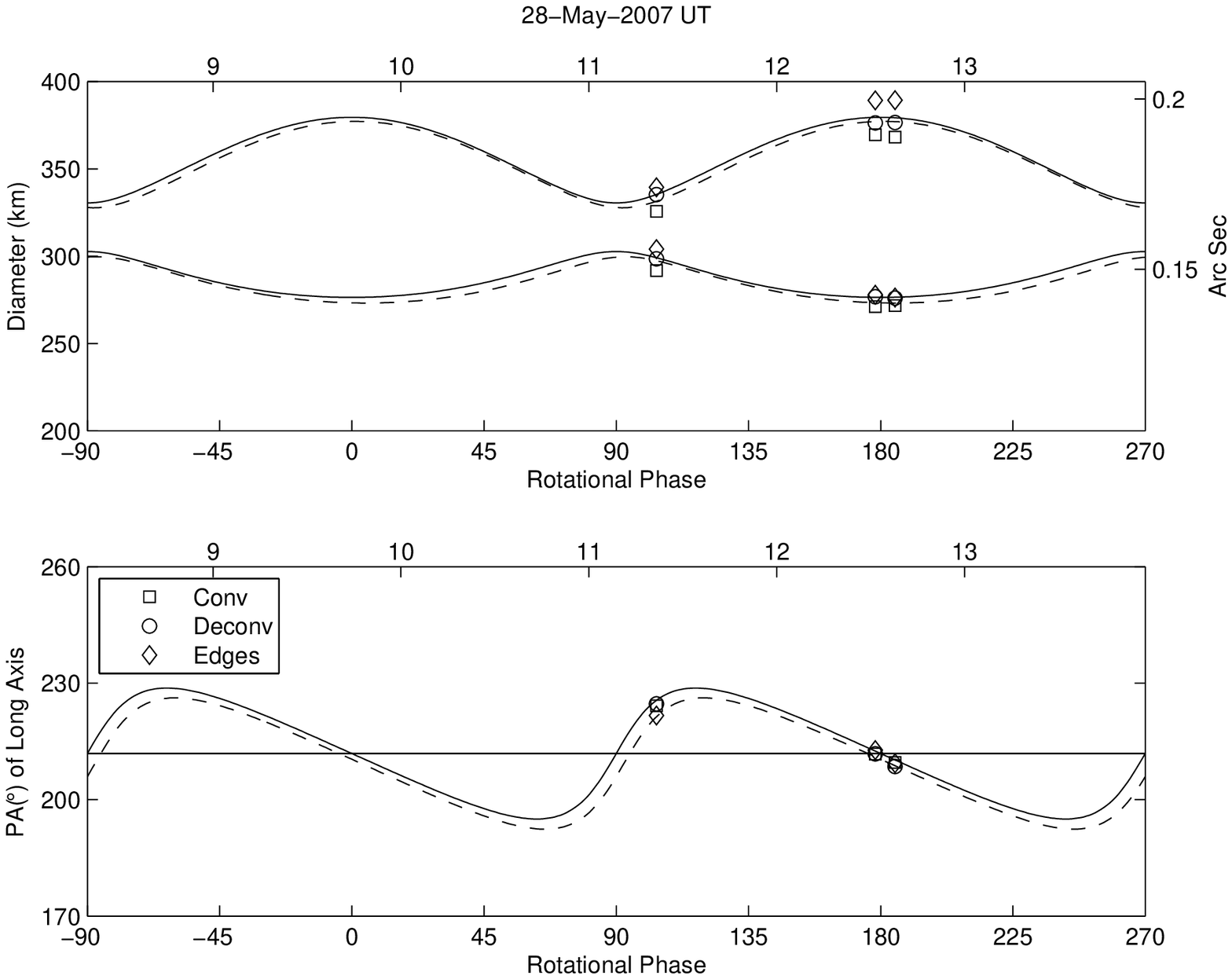}
  \caption{%
    Same as Fig.~\ref{fig: stea05}, but for 2007.
    The maximum that occurs
    at $9.74 \pm 0.01$, lighttime corrected, is the same hemisphere as the maximum
    that occurs at 10.30 UT in Fig.~\ref{fig: stea05}.
  \label{fig: stea07}
  }
\end{center}
\end{figure}

    \indent We also acquired AO observations of (52)\,Europa at Keck
     in 2007 (Table~\ref{tab: obs}).
    Following the recipe from the last section, we
    formed the mean apparent parameters
    from the three methods already described
    (PBD, deconvolved
    images, and outlines from the deconvolved images).
    Although not expected to
    yield significant results because the three 2007 observations provide only nine
    observables to find six unknowns, we nevertheless fit the three observations
    for a triaxial ellipsoid (Table~\ref{tab: stea07} and Fig.~\ref{fig: stea07}),
    and found that the model is in
    surprisingly good agreement with the results from the 2005 set in
    Table~\ref{tab: stea05}.
%
%%%%%%------ FIGURE --- Begin --- STEA Curve ------%%%%%%
%%%%%%------ FIGURE ---  End  --- STEA Curve ------%%%%%%
%
%
%
%%%%%%%------ TABLE --- Begin --- STEA Parameters ------%%%%%%
\begin{table}[!ht]
\begin{center}
  \begin{tabular}{cc}
    \hline\hline
    Parameter  & Mean \\
    \hline
    a (km)  & 379 $\pm$ \addr{1}\sout{2}  \\
    b (km)  & 330 $\pm$ \addr{1}\sout{3}  \\
    c (km)  & \addr{225}\sout{226} $\pm$ \addr{9}\sout{15} \\
    \hline
    SEP$_\beta$ (\degr)& -41 $\pm$ 5 \\
    PA$_\textrm{node}$ (\degr)& 212 $\pm$ 1 \\
    $\psi_0$ (UT) & 9.74 $\pm$ \addr{0.01}\sout{0.04} \\
    \hline
    EQJ2000 ($\alpha_0$,$\delta_0$ in \degr) &
    258;+11 \\
    $\sigma$ radius (\degr)  & \addr{1}\sout{2} \\
    ECJ2000 ($\lambda_0$,$\beta_0$ in \degr) &
    256;+34 \\
    \hline
  \end{tabular}
  \caption{
    Triaxial Ellipsoid Fit Parameters from 2007 observations.
    Uncertainties reported here are formal error bars of the fit, see
    the text for a discussion on the systematics.
    \label{tab: stea07}
  }
\end{center}
\end{table}
%%%%%%%------ TABLE ---  End  --- STEA Parameters ------%%%%%%

  \subsection{2003\,October\,12}
\begin{figure}[!t]
\begin{center}
  \includegraphics[width=.45\textwidth]{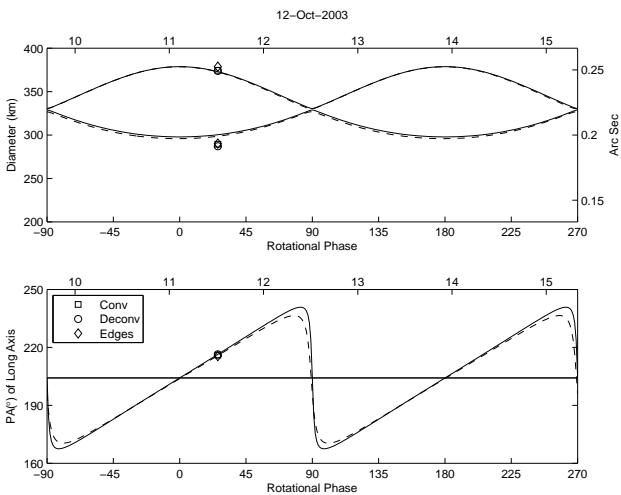}
  \caption{%
    Global fit and 2003 data.
  \label{fig: stea03}
  }
\end{center}
\end{figure}

    \indent The single set of AO images of (52)\,Europa taken in 2003
    (Table~\ref{tab: obs}) does not
    allow an independent fit for a triaxial solution because it only
    provides three 
    observables for six unknowns.
    We use these early Keck AO images, however, in 
    a global fit in the next section. 
    Fig.~\ref{fig: stea03} shows the global fit prediction for
    the 2003 epoch, together with those data.
%
%
%
%%%%%%------ FIGURE --- Begin --- STEA Curve ------%%%%%%
%%%%%%------ FIGURE ---  End  --- STEA Curve ------%%%%%%

  \subsection{A global solution for all epochs}
    \indent We can tie the 2003, 2005, and 2007 observations of (52)\,Europa together
    into one simultaneous global fit
    (Drummond et al., 2012, in preparation), using
    the sidereal period of P$_s$\,=\,0.2345816\,days
    (with an uncertainty of 2 in the last digit)
      derived by \citet{2004-AA-416-Michalowski}.
    Along with the global solution for the triaxial
    dimensions and pole in Table~\ref{tab: global},
    we list the three parameters that differ due
    to the changing angles for each date. \\
\indent  \addr{Statistical uncertainties for the dimensional parameters,
as well as those involving angles, such as pole position and 
    longitude of the node, come from the non-linear least-squares
fit for the 6 parameters
    that define the TE model, the 3 diameters and 3 Euler angles.}
    Systematic effects can arise
    in the process of constructing a 3-D description of an
    asteroid from information limited to a 2-D plane (images).
    Therefore, one needs to be particularly vigilant regarding model
    assumptions, and their appropriateness for a particular situation.
    While the uncertainties derived for the parameters as fit by the
    model are straightforward, estimating the systematic effects 
    that are present is not.
    Deriving realistic (and therefore, directly applicable by other
    workers) uncertainties for our results, including
    possible systematics, is the most challenging aspect of our work.

    We have carefully calibrated some of these uncertainties by making observations
    of external sources (\eg, the moons of Saturn) of known size.
    \addr{One of the results of that work has shown that our systematic
    uncertainties are larger for objects of smaller angular diameter,
    until we reach a limit (at about 0.09\arcsec for a 10\,m telescope)
    where we can no longer get reliable sizes.  Aspect ratios of
    projected shapes are still possible, but absolute sizes break down.
    We have found that our systematics from these tests span about
    1--4\% per linear dimension.   In addition, we have also
    imaged targets of spacecraft missions prior to flyby (see KOALA section).
    In the case of (21)\,Lutetia, despite an angular size of only 0.10\arcsec, our
    resulting models were good to 2\% in size and 2\,km RMS 
    in topography on a 100\,km object 
      \citep[see][]{2012-PSS-66-Carry}. } 

    We can also 
    compare our TE results with those of KOALA (see below),
    in cases where we have adequate observations. 
    In particular, we have such comparisons for four asteroids,
    including (52)\,Europa.
\addr{We can look for consistency, not
    only between the two techniques, but in sub-sets of data to
    learn how far we fall from the ``correct" values.  We can
    also compare the results of data sets from different 
    years.  Our upcoming article, mentioned above
    (Drummond et al., in preparation) will be a stand-alone treatment of
     the global fitting technique
     and calibration that will
    include much detail on uncertainties.}  
    \sout{(see Drummond et al., in preparation).} 
    For the present results, we have determined that we 
    should add quadratically systematic uncertainties of 4.1\%, \sout{2.4}\addr{2.3}\%, and 3.8\%
    to the \addr{TE-derived} fit errors (given in Table~\ref{tab: global}) \addr{for
$a,b,c$, respectively.}
\addr{The resulting
total uncertainty estimates for the $a,b,c$\, diameters are}   \sout{to the
    TE-derived ellipsoid diameters 
    of (52)\,Europa to be:} 16\,$\times$\,8\,$\times$\,10\,km\addr{, with
    a 7 degree systematic uncertainty for the orientation of the spin axis.}
    See Fig.~\ref{fig: globes} for a visualization of  the orientation
    of (52)\,Europa on the plane of the sky.

%
%
%%%%%%%------ TABLE --- Begin --- STEA Global Fit ------%%%%%%
\begin{table*}[!ht]
\begin{center}
  \begin{tabular}{cccccc}
    \hline\hline
    Diameter (km)  & Pole & Param & 2003\,Oct\,12 & 2005\,Jan\,20 & 2007\,May\,28 \\
    \hline
    a = 379 $\pm$ 2 & ($\alpha_0$,$\delta_0$) = 257\degr;+12\degr & SEP$_\beta$ (\degr)       &   +49 $\pm$ 1   &   +23 $\pm$ 1    &  -40 $\pm$ 1  \\
    b = 330 $\pm$ 2 & $\sigma$ radius         = 1\degr            & PA$_\textrm{node}$ (\degr)&   204 $\pm$ 1   &   339 $\pm$ 1    &  213 $\pm$ 1   \\
    c = 249 $\pm$ 3 & ($\lambda_0$,$\beta_0$) = 255\degr;+35\degr & $\psi_0$ (UT)            & 11.11 $\pm$ 0.02 & 10.31 $\pm$ 0.02 & 9.72 $\pm$ 0.02 \\
    \hline
  \end{tabular}
  \caption{
    Results for the global fit.
    Uncertainties reported here are formal error bars of the fit.
    Including systematic effects raises the total uncertainties to 
    16\,$\times$\,8\,$\times$\,10\,km for the three ellipsoid
    diameters,
    and to \sout{3}\addr{7}\degr in the pole.
    \label{tab: global}
  }
\end{center}
\end{table*}
%%%%%%%------ TABLE ---  End  --- STEA Global Fit ------%%%%%%

%%%%%%%%%%%%%%%%%%%%%%%%%%%%%%%%%%%%%%%%%%%%%%%%%%%%%%%%%%%%
%%%%%%    TAG     %%%%%%%%%%%    COMPARISON       %%%%%%%%%%
%%%%%%%%%%%%%%%%%%%%%%%%%%%%%%%%%%%%%%%%%%%%%%%%%%%%%%%%%%%%
\section{Comparison of (52)\,Europa to Lightcurves Inversion Model}
\begin{figure}[!t]
\begin{center}
  \includegraphics[width=.45\textwidth]{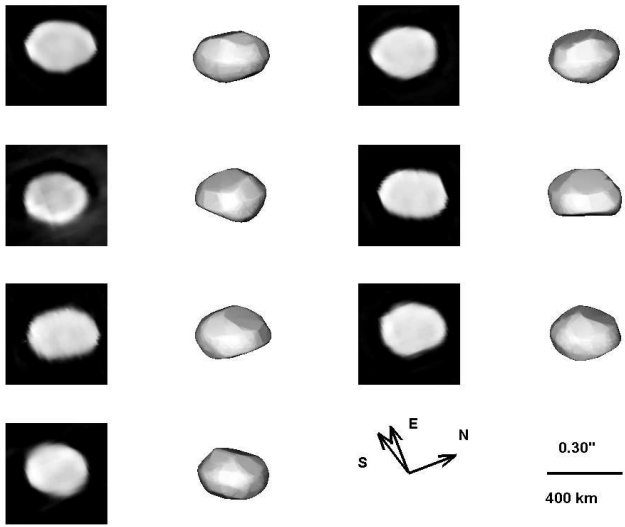}
  \caption{%
    Comparison of our (2005) deconvolved K images from
    Fig.~\ref{fig: deconv} (columns 1 and 3) with
    the web model of \citet{2004-AA-416-Michalowski}, projected forward
    from 1983 using their
    sidereal period of 0.2345816 days and an update (although nearly
    identical) to thier pole from DAMIT (columns 2 and 4).
  \label{fig: lc2005}
  }
\end{center}
\end{figure}

\begin{figure}[!t]
\begin{center}
  \includegraphics[width=.45\textwidth]{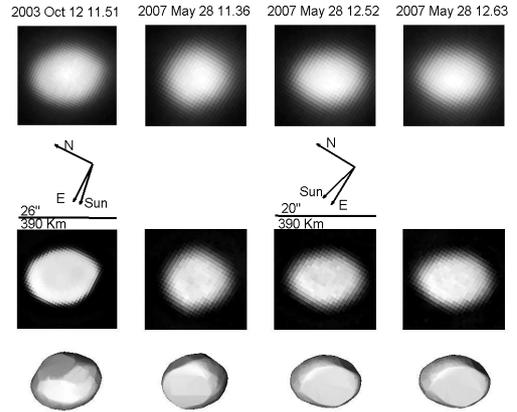}
  \caption{%
    Same as Fig.~\ref{fig: lc2005}, but for 2003 and 2007.
    In addition to the deconvolved images in the middle row, we show
    the non-deconvoled, shifted, and centered
    images in the top row for each epoch. In 2003, (52)\,Europa was 1.3 times closer
    than in 2007 resulting in different scales for the two years.
  \label{fig: lc_all}
  }
\end{center}
\end{figure}

  \indent From optical lightcurves of (52)\,Europa,
  \citet{2004-AA-416-Michalowski} found a
  rotational pole at [$\lambda_0$, $\beta_0$]=[252\degr, +38\degr],
  with a 5\degr~uncertainty in each Ecliptic
  coordinate. It is the pole closest to ours in
  Fig.~\ref{fig: pole}, about 6\degr~away.
  They derived an a/b axial ratio of 1.15, the same as our
  $1.15 \pm 0.04$,
  and a b/c ratio of 1.3, compared with our
  $1.33 \pm 0.05$.\\
  \indent Figure~\ref{fig: lc2005} is a side by side comparison
  of our deconvolved Kp images, from 2005\,January\,20
  (from Fig.~\ref{fig: deconv}) and the 
  \citeauthor{2004-AA-416-Michalowski} model,
  using the updated rotational pole
  for the model at [$\lambda_0$, $\beta_0$]=[251\degr; +35\degr] from \addr{the} DAMIT
  \citep{2010-AA-513-Durech} web
  site\footnote{\href{http://astro.troja.mff.cuni.cz/projects/asteroids3D/web.php}{http://astro.troja.mff.cuni.cz/projects/asteroids3D/web.php}}.
  Figure~\ref{fig: lc_all} show\addr{s} \sout{a} comparison between our convolved and deconvolved
  images and the lightcurves inversion model for 2003 and 2007. \\
  \indent The overall agreement between our AO deconvolved images and the
  model predictions is excellent.
  \sout{However, a} \addr{A} careful examination of
  Figs~\ref{fig: lc2005}
  and~\ref{fig: lc_all}, \addr{however,}
  will show
  \sout{that there are} edge features that are seen in one but not the
  other,
  requiring the development of an updated shape model, as
  discussed in following section.
  Despite these features, (52)\,Europa is still well-modeled as a smooth triaxial
  ellipsoid.

%
%
%
%%%%%%------ FIGURE --- Begin --- Comparison LC ------%%%%%%
%%%%%%------ FIGURE ---  End  --- Comparison LC ------%%%%%%
%
%
%%%%%%%------ FIGURE --- Begin --- Comparison LC ------%%%%%%
%%%%%%------ FIGURE ---  End  --- Comparison LC ------%%%%%%

\section{KOALA 3-D shape model\label{sec: koala}}
\begin{figure}[!t]
\begin{center}
  \includegraphics[width=.45\textwidth]{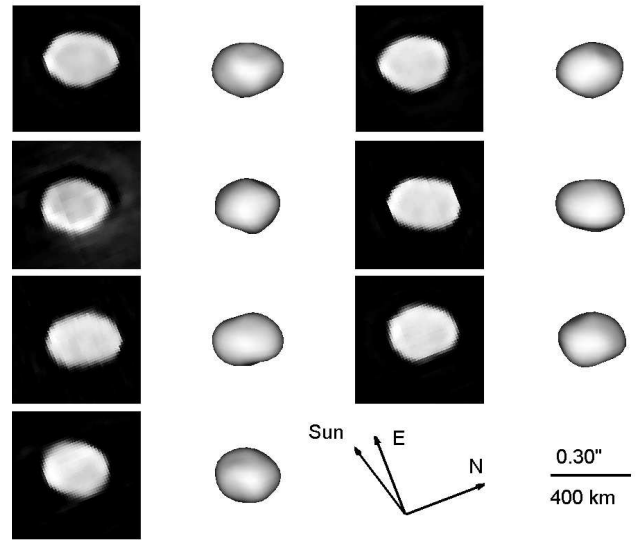}
  \caption{%
    Comparison of our 2005 deconvolved Kp images from
    Fig.~\ref{fig: deconv} (columns 1 and 3) with
    the KOALA model described here
  \label{fig: koala}
  }
\end{center}
\end{figure}

\begin{figure}[!t]
\begin{center}
  \includegraphics[width=.245\textwidth]{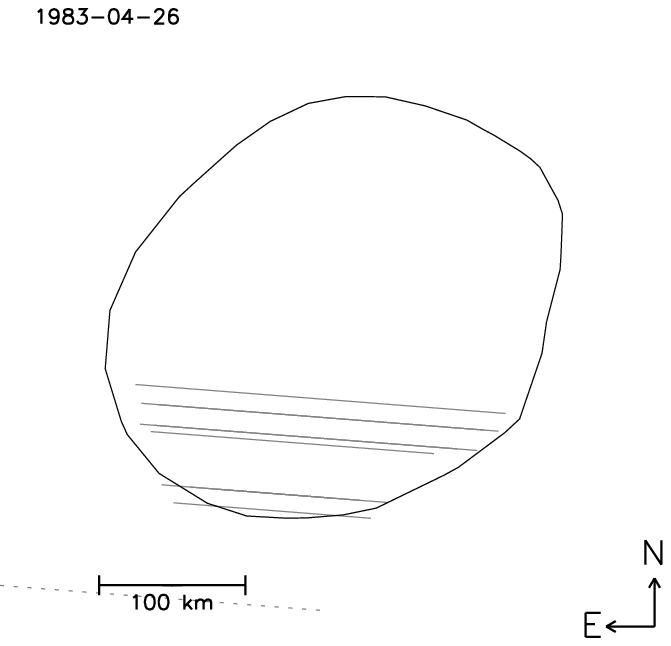}%
  \includegraphics[width=.245\textwidth]{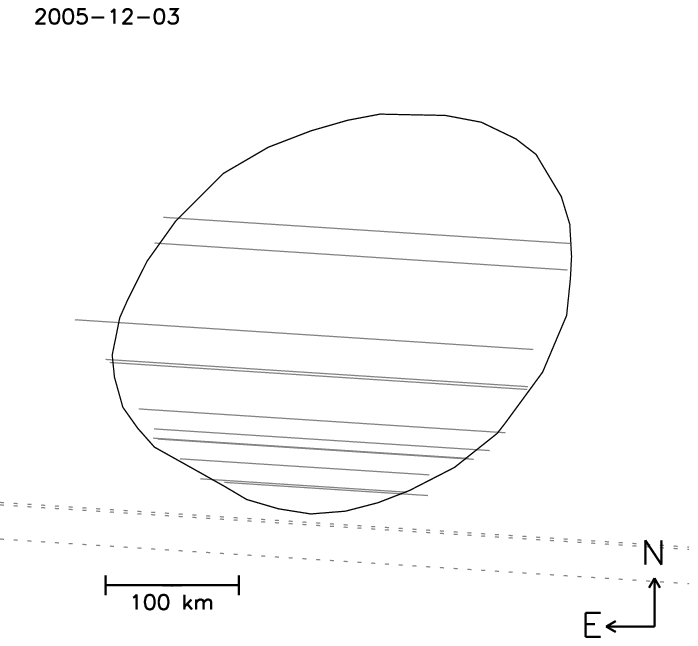}\\
  \includegraphics[width=.245\textwidth]{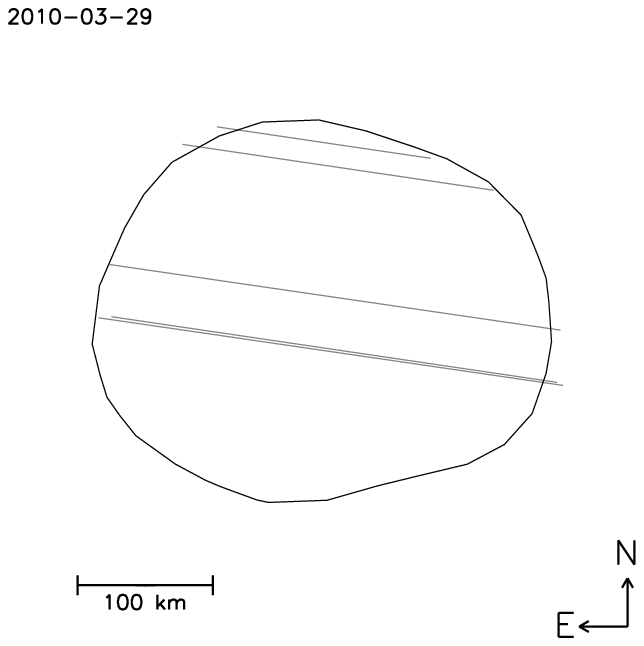}%
  \includegraphics[width=.245\textwidth]{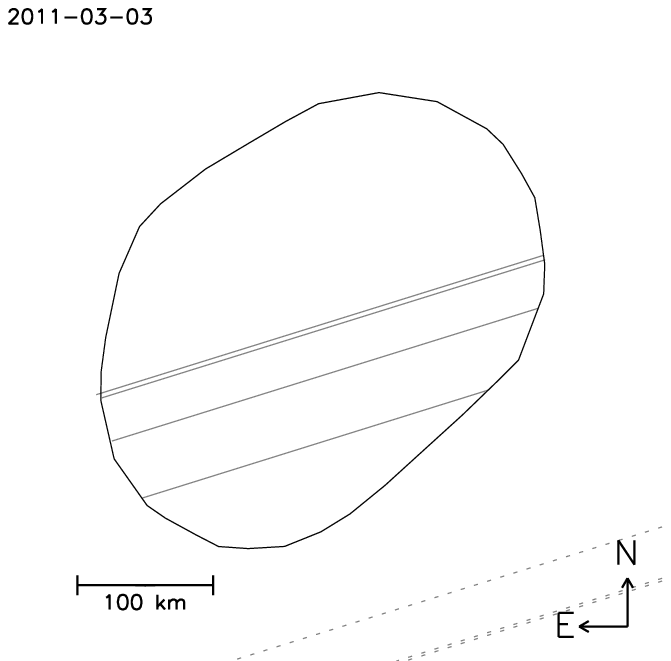}
  \caption{%
    Comparison of the four stellar occultations with the KOALA shape
    model. Solid and dashed grey lines represent positive (hits)
    and negative (misses) chords, respectively.
    Black contours are the projection of the KOALA
    3-D shape model on the plane of the sky at each occultation epoch.
  \label{fig: koalaocc}
  }
\end{center}
\end{figure}

  \indent We construct a 3-D shape model of (52)\,Europa to give a better rendering
  of the apparent shape visible in the images.
  For that, we use our KOALA algorithm
  \citep{2010-Icarus-205-Carry-a, 2011-IPI-5-Kaasalainen}
  that makes combined use of optical lightcurves, stellar
  occultations timings, and profiles from disk-resolved images.
  The results of KOALA have been recently validated at (21)\,Lutetia
  by the images taken by the ESA Rosetta mission:
  The 3-D shape model and spin orientation
  determined before the encounter by combining AO images and
  lightcurves \citep{2010-AA-523-Carry}
  were in complete agreement with images and results from
  the flyby \citep{2011-Science-334-Sierks, 2012-PSS-66-Carry}.
  Axial dimensions from KOALA were determined within
  2\% of the the actual values and RMS differences in topography
  were only 2\,km. \\
  \indent We use here the 18 imaging epochs described in
  Sect.~\ref{sec: obs}, together with 49 lightcurves taken
  between 1979 and 2011
  \citep[we acquired 8 additional lightcurves within the
  CdR/CdL collaboration with respect to the 41 lightcurves presented
  by][]{2004-AA-416-Michalowski}, and 4 stellar occultations
  \citep[timings taken from][]{PDSSBN-OCC}.
  A comparison of the KOALA 3-D shape model with the AO images from
  2005 is presented in Fig.~\ref{fig: koala}.
  The agreement between the 3-D shape model and the data is very good.
  The typical deviation with the 18 imaging contours is
  of 0.2\,pixel, corresponding to a few km.
  The 49 lightcurves are rendered at a level of 0.03 mag, \ie, close
  to the intrinsic level of uncertainty of the data.
  Finally, the residuals between the occultation chords and the model
  are 13\,km, on average, mainly owing to the lower quality of 1983
  occultation timings (residuals of 19\,km, compared to 11, 13, and
  6\,km for the other epochs).
  Figure~\ref{fig: koalaocc} shows these chords mapped onto the
  projections of the 3-D KOALA model for the epochs of the
  occultations. \\
  \indent The 3-D shape derived with KOALA is
  close to an ellipsoid, validating (52)\,Europa as a Standard Triaxial
  Ellipsoid Asteroid
  \citep[STEA,][]{2008-DPS-40-Drummond}.
  Fitting the KOALA model as a triaxial ellipsoid yields diameters
  of \sout{371}\addr{368}\,$\times$\,327\,$\times$\,\sout{256}\addr{255}\,km, in excellent agreement 
  with the diameters and total uncertainties in Table~\ref{tab:
    global}.
  The volume-equivalent spherical-diameter of the KOALA 3-D shape model\addr{, derived
by summing volume cells,}  is
    312\,$\pm$\,6\,km, in excellent agreement with the TE analysis
    presented above.
\addr{The KOALA model yields a spin pole  within 3\degr of
[$\lambda_0$, $\beta_0$]=[254\degr; +37\degr] or
[$\alpha_0$, $\delta_0$]=[257\degr; +15\degr], also
close to the TE result.}
    The shape model can be downloaded \sout{on} \addr{from the} DAMIT web
    page.

%
%
%
%
%\section{Comparison with Davida\label{sec: comparison}}
\section{Occurrence of large facets on C-type asteroids\label{sec: comparison}}

  The 3-D shape model presents two broad shallow depressions, \addr{probably
  best noted in the lower right of Fig. 12.  They can also be seen on the
  tops and bottoms of the asteroid images in column 1, panel 3, and column 3, panel 2.}
  The departures from an ellipsoid, however, are not nearly as
  significant as the apparent giant facets seen in our analysis of
  (511)\,Davida \citep{2007-Icarus-191-Conrad}, nor as  
  prominent, relative to body size, as the giant craters seen on
  (253)\,Mathilde \citep{1997-Science-278-Veverka}. 
  We chose Mathilde as a prototypical object displaying giant features
  seen in profile (craters/facets), although
  Mathilde was a much smaller
  asteroid than Davida. But it turns out that (52)\,Europa is almost a twin
  of Davida in many respects: both are C-type asteroids 
  of very nearly the same size, they have similar spin periods, and
  they have similar 
  orbital properties, so they have likely seen the same impact
  environment (although Davida does 
  have a bit larger eccenticity and inclination).
  In the Davida paper, 
  we went to some length to demonstrate that Davida could have
  encountered impacts of 
  the size necessary to produce the giant facets seen, without having
  broken up the 
  body.  So given the similarities between Davida and (52)\,Europa, one
  might now wonder 
  how likely it is that (52)\,Europa would \textit{not} show such facets (or at
  least not show facets that are quite as prominent). 

  Returning to our analysis in the Davida paper, we
  estimated that Davida should have had about 2.5 impacts large enough to
  make such a giant crater during its lifetime. This led to the conclusion that if the
  facets seen were indeed craters, seen edge-on, as on Mathilde, they would not
  be unexpected. The same statistics should hold true for (52)\,Europa. But with an
  expected total of only 2.5 impacts of this size during its lifetime, the chances are
  also reasonable that it did not encounter any such impacts. We therefore
  conclude that \textit{not seeing} such prominent features on a twin
  such as (52)\,Europa could also be expected.
  Of course, the flux of smaller impactors would be higher, and
  these would be responsible for the perhaps less prominent edge features
  that we do see.  \addr{Given that the viewing geometry has to be just 
  right to see these types of facets, it is possible that observational
circumstances have
   conspired such that we missed some giant feature, or such that those
   facets we do see are less pronounced or are particularly hard to follow
   with rotation. We have a fairly
   wide range of latitudes and longitudes in our data set, however, so
   the chances of missing something as prominent as a Davida-style facet
   are diminished, and we assert that Europa appears qualitatively
   different than Davida.}

%%%%%%------ FIGURE --- Begin --- Comparison KOALA ------%%%%%%
%%%%%%------ FIGURE ---  End  --- Comparison KOALA ------%%%%%%
%

%%%%%%------ FIGURE --- Begin --- Comparison KOALA ------%%%%%%
%%%%%%------ FIGURE ---  End  --- Comparison KOALA ------%%%%%%
%

%
%
%%%%%%%%%%%%%%%%%%%%%%%%%%%%%%%%%%%%%%%%%%%%%%%%%%%%%%%%%%%%%%%
%%%%%%    TAG     %%%%%%%%%%%   PHYSICAL PROPERTIES    %%%%%%%%
%%%%%%%%%%%%%%%%%%%%%%%%%%%%%%%%%%%%%%%%%%%%%%%%%%%%%%%%%%%%%%%
\section{Density of (52)\,Europa}
  \indent There are 17 estimates of the mass of (52)\,Europa
  available in the literature, derived either from the analysis of the
  orbit's deflection during close approaches of minor planets to
  (52)\,Europa \citep[\eg,][]{2001-AA-374-Michalak},
  or from a general adjustment of the parameters used to generate
  the ephemeris of the planets and
  asteroids in the Solar System
  \citep[\eg,][]{2009-AA-507-Fienga}.
  We adopt here the weighted mean of these determinations
  \citep[following the selection discussed in][]{2012-PSS-Carry}:
  $(2.38$\,$\pm$\,$0.58) \times 10^{19}$\,kg. \\
  \indent \addr{In general, the differences in volume between the
triaxial and the KOALA models are small.  Here, that difference
is less than 1\%, which
would lead to a volume difference of less than 3\%.  When assigning
uncertainties to our sizes (from either method), we not only
assess the derivable statistical uncertainties, but we must
also provide an estimate of systematic effects, of which this
difference is an example.  The uncertainties used already
include potential differences between the models. 
Because of the added
topographic detail provided by the KOALA model, we choose,
in this case, to
use the KOALA-derived volume of
  $(1.59$\,$\pm$\,0.10) $\times 10^{7}$\,km$^{3}$,}
\sout{Given the volume of
  $(1.63$\,$\pm$\,0.10) $\times 10^{7}$\,km$^{3}$
  that we derive here, we find} \addr{giving}
a density of
  1.5\,$\pm$\,0.4\,g\,cm$^{-3}$.
  This bulk density falls within the 
  observed range of densities for C-type asteroids.
  Here, the uncertainty is mainly due to the uncertainty on 
  the mass determination (24\%) rather than the volume uncertainty
  of 6\%.
  Thus, we are at the point in the study of the density of asteroids
  where the uncertainty on the volume is no longer the limiting
  factor
  \citep[volume determination remains generally the limiting factor
    when the mass is estimated from a spacecraft encounter or a
    satellite, see the review by][]{2012-PSS-Carry}.

  Dedicated observing programs and
  theoretical work are now needed to derive more accurate masses of
  large main-belt asteroids. The advent of the Gaia mission
  (expected launch 2013) should contribute a large number of new, improved masses
  \citep[see][for instance]{2007-AA-472-Mouret}.
  With these more reliable volumes and masses, we can derive improved
\addr{densities and porosities,}
\sout{  density and porosity estimates,} which in turn 
  will allow us to better understand how density and porosity may be
  related to taxonomic class, absolute diameters, or location (\eg,
  inner vs. outer main belt). And this 
  highlights the importance of continuing to push for more AO
  observations of asteroids for size/shape determination, from the 
  best facilities, and the continued development of techniques, such
  as KOALA, that combine multiple data types (hopefully, 
  eventually to include thermal radiometry and radar echoes).

%
%
%%%%%%%%%%%%%%%%%%%%%%%%%%%%%%%%%%%%%%%%%%%%%%%%%%%%%%%%%%%%
%%%%%%    TAG     %%%%%%%%%%%    CONCLUSIONS      %%%%%%%%%%
%%%%%%%%%%%%%%%%%%%%%%%%%%%%%%%%%%%%%%%%%%%%%%%%%%%%%%%%%%%%
\section{Summary}
  \indent At this point, (52)\,Europa can be considered for
    membership as a Standard Triaxial Ellipsoid Asteroid
%  \citep[STEAs, see][]{2009-Icarus-202-Drummond} 
  \citep[STEAs, see][]{2008-DPS-40-Drummond} 
  because it
  is so well modeled as an ellipsoid
  \citep[like asteroid (511)\,Davida, see][]{2007-Icarus-191-Conrad}.
  The ellipses projected by these standard ellipsoids
  can be predicted well into the future or past, and therefore, can be used as
  calibration objects for other techniques used in studying asteroids.
  \citet{2007-Icarus-191-Conrad} and Drummond et al. (in preparation)
  detail the equations necessary to
  predict the asterocentric latitudes and longitudes, and
  \citet{2000-LGSAO-Drummond}
  show how to derive the projected ellipse parameters from the asterocentric
  latitudes and longitudes. For example, (52)\,Europa's asterocentric latitude can
  be predicted to within the error of its rotational pole,
  \sout{3}\addr{7}\degr,
  and its asterocentric
  longitude to within 0.5\degr/yr since the date of the
  most recent epoch reported here (2007\,May\,28).
  The longitude uncertainty arises from the formal uncertainty in the
  sidereal period, but in fact, judging by the good agreement shown between
  the images and lightcurves inversion model projected forward from 1983, 
  longitudes should be predictable to a much higher
    accuracy than these values indicate.
  The projected major or minor axis dimensions can
  be predicted to within approximately the uncertainty found here of
  5--10\,km,
  and the orientation of the apparent ellipse to within
  \sout{5}\addr{2}\degr. \\
  \indent We are fortunate to have both the
  \addr{triaxial ellipsoid \citep[TE,][]{2009-Icarus-202-Drummond}
    and the KOALA 
    \citep{2010-Icarus-205-Carry-a}}
    techniques
  available for our analysis of AO images of asteroids.  Each has its own
  strengths. TE requires relatively few images, can return shape/size/pole
  information amazingly quickly, is generally insenstive to changes in
  the PSF, and is usually adequate to get the basic asteroid
  parameters. For more detailed 3-D shape information we can rely on
  KOALA. \addr{Unlike lightcurve inversion alone, KOALA can obtain absolute
sizes, and is sensitive to concavities.}
The methods can be used to validate each other, as
  we found exceedingly useful during our analysis of the Lutetia data, 
  prior to the Rosetta flyby
  \citep{2010-AA-523-Drummond, 2010-AA-523-Carry}. 
  And while a detailed 3-D shape model might be seen to
  supercede the triaxial assumption of TE, that is not necessarily the
  case.  As an example, our AO imaging of the close flyby of Near-Earth
  Asteroid 2005~YU55 from Keck, in November 2011, resulted in
  almost immediate size and shape information from TE
  \citep{2011-IAUC-9242-Merline}.
  In futher analysis, we had hoped to use numerous lightcurves,
  taken near the time of the flyby, to help refine the size/shape with
  KOALA.  But despite our efforts, the lightcurve information
   \addr{on 2005~YU55}
  so-far is insufficient
  (mostly due to a very slow spin period)
  to allow KOALA to improve significantly on
  TE. \addr{This demonstrates the importance of having both
  methods available for analysis of our asteroid data.}
   \sout{We therefore intend to employ both methods in
  analysis of our data.} \\
  \indent New imaging, lightcurve, and occultation data will be added to
  our overall analysis for (52)\,Europa as they become available.  These may allow us to
  distinguish whether any of the somewhat-flattened edges seen on
  (52)\,Europa in our existing data sets are indeed facets or craters of the
  type seen on Davida and Mathilde, and to better evaluate the
  extent and morphology of any departure from a pure ellipsoid.
  The techniques we are
  developing here (both observational and in data analysis) will
  allow us to make immediate and substantial
  advances once data from new, larger telescopes can be acquired.

%%%%%%%%%%%%%%%%%%%%%%%%%%%%%%%%%%%%%%%%%%%%%%%%%%%%%%%%%%%%
%%%%%%    TAG     %%%%%%%%%%%    ACK              %%%%%%%%%%
%%%%%%%%%%%%%%%%%%%%%%%%%%%%%%%%%%%%%%%%%%%%%%%%%%%%%%%%%%%%
\section*{Acknowledgments}
  \indent This work presented here was supported by grants from NASA's Planetary
  Astronomy Program and the U.S. National Science Foundation, Planetary
  Astronomy Program.   We are grateful for telescope time made available to us
  by the NASA TAC, and also for the support of our collaborators on Team Keck,
  the Keck science staff. 
  The work of J. {\v D} was supported by grant P209/10/0537
  of the Czech Science Foundation and by the Research Program 
  MSM0021620860 of the Ministry of Education. \\
  \indent This research has made use of NASA's Astrophysics Data System.
  The authors wish to recognize and acknowledge the very significant
  cultural role and reverence that the summit of Mauna Kea has always
  had within the indigenous Hawaiian community.  We are most fortunate
  to have the opportunity to conduct observations from this mountain.


\bibliographystyle{icaruslike}
%\bibliographystyle{elsarticle-harv}
%\bibliography{biblio}

\end{document}